%% file: main.tex
\documentclass[10pt,twocolumn,letterpaper]{article}

\usepackage{cvpr}              % To produce the CAMERA-READY version

\definecolor{cvprblue}{rgb}{0.21,0.49,0.74}
\usepackage[pagebackref,breaklinks,colorlinks,allcolors=cvprblue]{hyperref}

\def\paperID{25} % *** Enter the Paper ID here
\def\confName{CVPR}
\def\confYear{2025}

\title{Zero-Shot Denoising for Fluorescence Lifetime Imaging Microscopy with Intensity-Guided Learning}

\author{
Hao Chen, Julian Najera, Dagmawit Geresu, Meenal Datta, Cody Smith, Scott Howard \\
University of Notre Dame, Notre Dame, IN, USA \\
{\tt\small \{hchen27, jnajera2, dgeresu, mdatta, csmith67, showard\}@nd.edu}
}
\begin{document}
\maketitle
\input{sec/0_abstract}    
\input{sec/1_intro}
\input{sec/2_related_work}
\input{sec/3_method}
\input{sec/4_experiments}
\input{sec/5_conclusions}
{
    \small
    \bibliographystyle{ieeenat_fullname}
    \bibliography{main}
}


\end{document}

%% file: sec/0_abstract.tex
\begin{abstract}
Multimodal and multi-information microscopy techniques such as Fluorescence Lifetime Imaging Microscopy (FLIM) extend the informational channels beyond intensity-based fluorescence microscopy but suffer from reduced image quality due to complex noise patterns. In FLIM, the intrinsic relationship between intensity and lifetime information means noise in each channel exhibits a multivariate dependence across channels without necessarily sharing structural features. Based on this, we present a novel Zero-Shot Denoising Framework with an Intensity-Guided Learning approach. Our correlation-preserving strategy maintains important biological information that might be lost when channels are processed independently. Our framework implements separate processing paths for each channel and utilizes a pre-trained intensity denoising prior to guide the refinement of lifetime components across multiple channels.  Through experiments on real-world FLIM-acquired biological samples, we show that our approach outperforms existing methods in both noise reduction and lifetime preservation, providing more reliable extraction of physiological and molecular information.
\end{abstract}

%% file: sec/1_intro.tex
\section{Introduction}
\label{sec:intro}
Fluorescence microscopy techniques like confocal and two-photon microscopy have been widely utilized for biological imaging and analysis~\cite{sanderson2014fluorescence}. Building upon standard fluorescence microscopy, multimodal and multi-information microscopy provides additional data dimensions including lifetime, phase differences, and other parameters~\cite{Zhang21Instant,park2018quantitative,jares2003fret}. Among these, Fluorescence Lifetime Imaging Microscopy (FLIM) quantifies the exponential decay rate of fluorophore emission following excitation as lifetime. This technique acquires multi-channel data that simultaneously capture both lifetime and intensity distributions, providing critical insights into molecular microenvironments. The additional channels from lifetime enable measurement of biological processes independent of fluorophore concentration, making FLIM invaluable for applications ranging from cancer diagnostics to metabolic imaging~\cite{datta2020fluorescence,becker2012fluorescence}.

Despite these advantages, FLIM imaging quality is fundamentally limited by the quantum yield rate from biological samples~\cite{lippincott2001studying}. This low photon count introduces shot noise that distorts accurate lifetime measurements, especially when measuring molecular interactions between and within cells. Moreover, unlike conventional fluorescence microscopy, FLIM links lifetime and intensity channels simultaneously, meaning noise in one channel is a function of multiple channels. However structural information is not shared between different modalities, suggesting that denoising methods developed for one modality may not be directly applicable to another.

While traditional denoising algorithms and deep learning techniques have significantly improved image denoising performance \cite{Zhang17, ronneberger2015u}, adapting these methods to FLIM's multi-channel data is still challenging. Collecting paired noisy–clean FLIM training sets is particularly hard due to photobleaching effects and sample damage under repeated excitation, and limited photon counts further restrict the effective training of deep learning models.

Additionally, the diversity of sample types and fluorophores has resulted in a limited availability of FLIM-dedicated datasets, making new methods harder to develop. Although self-supervised approaches such as Noise2Void \cite{Krull19} address some of these issues, they often struggle with the spatially correlated noise patterns characteristic of FLIM imaging. Although multi-modal and multi-information microscopy techniques face limited benchmarks with most existing datasets being sample-specific, intensity-based contrast images are commonly available and widely used across these fields. Based on these observations, we present a novel Zero-Shot Denoising Framework with an Intensity-Guided Learning approach in FLIM to enable effective denoising without requiring paired training data for all FLIM channels, while preserving inherent correlations to maintain biologically meaningful data that would be lost through independent channel processing. 

The main contributions of this paper are:
\begin{itemize}
\item A zero-shot framework for denoising without paired ground truth data using intensity-lifetime correlations to preserve their physical relationships by applying an intensity denoising model as a structural prior.
\item A comprehensive loss function design with intensity guidance and other channel-wise constraints to ensure effective denoising while preserving lifetime information.
\item Extensive evaluation on real-world FLIM datasets, demonstrating superior performance in noise reduction and lifetime preservation compared to existing methods.
\end{itemize}

While our method is specifically developed for FLIM denoising, its core principles and methodological framework are broadly applicable to other multimodal microscopy systems, extending well beyond denoising tasks. The presented approach represents a meaningful advancement toward addressing the wider challenge of processing multimodal microscopy data acquired under low-photon conditions. Moreover, our technique could be effectively leveraged in intensity-based multispectral imaging modalities, as well as frequency-domain imaging and signal processing applications.

%% file: sec/2_related_work.tex
\section{Related Work}
\label{sec:related_work}
In this section, we review prior work on microscopy image denoising methods, covering both traditional and deep learning approaches with a focus on techniques designed for realistic microscopy applications. We also discuss the challenges presented by multi-channel and multimodal microscopy methods like FLIM, an area with very limited existing research. Lastly, we explore zero-shot learning methods, highlighting their potential for microscopy denoising, particularly in cases where limited labeled training data limits the applicability of supervised approaches, and examine how these methodologies could specifically enhance FLIM denoising performance.

\subsection{Image Denoising for Microscopy}
Microscopy images are degraded by severe noise resulting from physical and instrumental limitations inherent to imaging systems. Factors such as low photon counts, sample-induced distortions, electronic sensor noise, and photobleaching collectively reduce image quality~\cite{von2012fluorescence,klonis2002fluorescence,lippincott2001studying}. 

Numerous denoising algorithms have been developed to address these challenges, significantly improving microscopy image quality and enabling the extraction of valuable biological information.

\noindent \textbf{Traditional Denoising Approaches.} Like many other signal processing techniques, early microscopy denoising methods were adapted from approaches developed for natural image processing. Classical approaches such as Gaussian smoothing, median filtering, and wavelet thresholding~\cite{cristobal1996wavelet} have been widely applied to suppress noise in microscopy images. However, these methods often oversmooth fine cellular structures, resulting in loss of critical biological details. More sophisticated approaches such as Non-local Means (NLM)~\cite{buades2011non} and Block-Matching 3D (BM3D)~\cite{dabov2006image} exploit self-similarity patterns in images to perform collaborative noise averaging. While these methods work well for structured images, they struggle with biological data where mixed Poisson-Gaussian noise fundamentally limits imaging performance~\cite{sarder2006deconvolution,zhang2019poisson}. Total variation minimization~\cite{vogel1996iterative} and feature-based likelihood optimization denoising models~\cite{maji2019feature,foi2008practical} have been developed for microscopy image restoration. Although effective in certain contexts, these methods are computationally intensive and require careful parameter tuning. Optimization-based approaches that incorporate domain knowledge through specialized regularization terms~\cite{arigovindan2013high} have shown promise for specific microscopy modalities but often fail to generalize across diverse imaging conditions.

\noindent \textbf{Deep Learning-Based Denoising.} Recent deep learning-based image denoising has improved performance by exploiting large-scale datasets and crafted network architectures. Convolutional neural network (CNN)-based approaches like DnCNN~\cite{Zhang17} and U-Net variants~\cite{ronneberger2015u} have been widely adopted in microscopy image denoising. These models achieve state-of-the-art results by learning noise patterns from paired noisy-clean image datasets. However, due to the fundamental limitations of microscopy imaging across different systems and biological samples, obtaining comprehensive training datasets remains challenging or often impractical in the microscopy field. 

Self-training denoising including Noise2Noise~\cite{Lehtinen18}, Noise2Void~\cite{Krull19}, and Noise2Self~\cite{Batson19} effectively denoise microscopy images without noisy-clean pairs. However, their assumption of statistically independent, zero-mean noise fails to fit the biological imaging, where mixed Poisson-Gaussian noise and spatially correlated artifacts matter~\cite{sarder2006deconvolution, meiniel2018denoising}. Consequently, the application of deep learning-based denoising techniques to microscopy remains challenging due to the inherent diversity of imaging modalities and the complex, specimen-specific noise characteristics.

\noindent \textbf{Challenges in Multimodal and Multi-Information Microscopy.} 
Additional modalities in microscopy encode information beyond conventional intensity measurements, including exponential decay lifetime dynamics~\cite{becker2012fluorescence}, phase shift information~\cite{nguyen2022quantitative}, and polarization-resolved molecular orientation~\cite{levitt2009fluorescence}. While deep learning has significantly advanced microscopy denoising techniques, multimodal and multi-information microscopy methods such as FLIM present distinct challenges. 

In FLIM, fluorescence lifetime measurements are fundamentally constrained by the signal-to-noise ratio~\cite{zhang2016investigation,philip2003theoretical} and the complex distributed photon arrival statistics across different detection channels~\cite{digman2008phasor, gratton2003fluorescence}. Furthermore, the difficulty in obtaining paired noisy-clean training data in FLIM has limited the widespread application of deep learning methods in this field. The very limited denoising work on FLIM comes from DnCNN pre-trained models which directly apply to the lifetime phasor channel of FLIM~\cite{mannam2023improving}. The complex, modality-specific noise characteristics in multimodal microscopy data such as FLIM significantly limit the effectiveness of conventional denoising approaches.

\subsection{Zero-Shot Learning for Denoising}
Zero-shot learning represents a machine learning approach that enables models to recognize or process previously unseen classes during inference~\cite{wang2019survey, palatucci2009zero}. In the context of image denoising, zero-shot approaches allow algorithms to effectively remove noise patterns not encountered during the training phase. However, most self-supervised methods and techniques based on deep image prior~\cite{ulyanov2018deep} fail to achieve true zero-shot capabilities. These models still learn similar image statistics across the entire dataset, constraining their ability to generalize to completely novel noise distributions.

Recent denoising techniques such as Noise2Fast~\cite{lequyer2022fast} and Neighbor2Neighbor~\cite{huang2021neighbor2neighbor} have made significant progress toward achieving dataset-free denoising. Building upon these approaches, methods like Zero-Shot Noise2Noise~\cite{mansour2023zero} demonstrate the capability to generalize across novel noise distributions without requiring additional training data, instead utilizing internal image pair information for effective denoising.

Despite the promising results of zero-shot denoising approaches in addressing standard microscope noise effects, these methods demonstrate significant limitations when applied to multi-channel correlated microscopy images such as FLIM. The complex correlations between channels in these specialized imaging modalities prevent conventional zero-shot approaches from effectively learning the internal information necessary for accurate denoising. 

Nevertheless, zero-shot approaches remain valuable when considering their data-free training methodology. In microscopy contexts where multiple modality information exists, combining zero-shot techniques with traditional denoising methods on the intensity channel could provide effective guidance for enhanced results. However, the potential of such combined approaches in preserving critical physiological and molecular information encoded in FLIM signals remains largely unexplored. This gap represents an important direction for applications requiring high fidelity in functional imaging of biomarkers.

%% file: sec/3_method.tex
\section{Method}
\label{sec:method}

\begin{figure*}[t]
    \centering
    \includegraphics[width=0.95\textwidth]{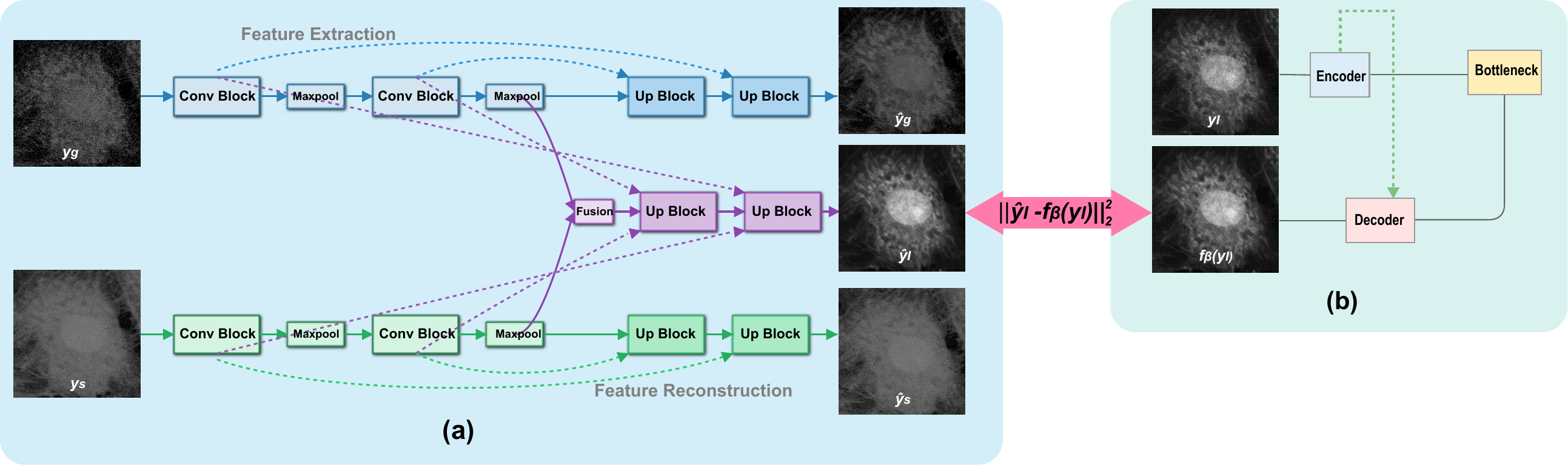}
    \caption{Architecture of our zero-shot FLIM denoising framework. (a) The dual-encoder network employs channel-specific encoders to process noisy input data, separately handling intensity-scaled parameters $y_g$ and $y_s$. Three specialized decoders then generate cleaned outputs: intensity-scaled $\hat{y_g}$, intensity $\hat{y_I}$ and intensity-scaled $\hat{y_s}$. (b) The pre-trained model utilizes a standard U-Net architecture that was trained on general fluorescence microscopy intensity data $y_I$ as a structural prior that guides the main network's denoising process.}
    \label{fig:architecture}
\end{figure*}

In this section, we introduce our Zero-Shot Denoising Framework with Intensity-Guided Learning. Our framework effectively reduces noise in fluorescence lifetime images across all channels, preserving important fluorescence lifetime details and enhancing image quality. In contrast to conventional deep learning denoising methods that typically rely on paired noisy-clean datasets for each image channel, the proposed method enables denoising in a zero-shot setting by exploiting the inherent correlation between fluorescence intensity images and their corresponding lifetime channels. We first formalize the problem and then describe the proposed network architecture. Afterward, we explain the training procedure and the strategy used for denoising.

\subsection{Problem Formulation}
\label{subsec:problem}

FLIM captures temporal fluorescence decay profiles to extract lifetime values $\tau$, which are quantified through the phasor coordinates~\cite{digman2008phasor}:
\begin{equation}
    g = \frac{1}{1+(\omega\tau)^2}, \quad s = \frac{\omega\tau}{1+(\omega\tau)^2},
\end{equation}
where $\omega$ represents the angular modulation frequency. The low photon counts and detection in FLIM acquisition introduce photon shot noise and additive Gaussian noise. For simplicity, a noisy FLIM image can be represented as $\mathbf{y} = (y_g, y_s)$, where:
\begin{equation}
    y_g = g \cdot I + n_g, \quad y_s = s \cdot I + n_s,
\end{equation}
$y_g$ and $y_s$ are the intensity-scaled noisy phasor coordinates, $I$ represents the clean fluorescence intensity signal, and $n_g$, $n_s$ denote the respective noise components. 

The goal of FLIM denoising is to estimate the clean signal vector $\hat{\mathbf{y}} = (\hat{y}_g, \hat{y}_s, \hat{y}_I)$, where $\hat{y}_g = \hat{I}\cdot\hat{g}$ and $\hat{y}_s = \hat{I}\cdot\hat{s}$. Here, $\hat{g}$ and $\hat{s}$ represent the estimated phasor coordinates, and $\hat{I}$ is the estimated fluorescence intensity map. This estimated vector $\hat{\mathbf{y}}$ should closely approximate the clean signal $\mathbf{y^*} = (g\cdot I, s\cdot I, I)$, while preserving the accuracy of the lifetime information embedded in the relationship between $\hat{g}$ and $\hat{s}$.

Unlike mixed Poisson-Gaussian noise datasets for confocal and two-photon microscopy, FLIM imaging lacks paired training data comprising noisy images and their clean reference. To overcome this constraint, we make use of a neural network that learns a noise-adaptive mapping function $f_\theta$ operating directly on multi-channel noisy FLIM acquisitions in a zero-shot manner to obtain clean vector estimations $\hat{\mathbf{y}}$ :
\begin{equation}
   \hat{\mathbf{y}} = f_\theta(\mathbf{y}, f_\beta(y_I)),
\end{equation}
where $y_I$ is the noisy intensity channel, $f_\beta(y_I)$ rrepresents the output of the pre-trained intensity denoising model $f_\beta$ trained on mixed Poisson-Gaussian datasets~\cite{zhang2019poisson}, which serves as a structural prior to guide the lifetime components denoising process.

\subsection{Network Architecture}
\label{subsec:architecture}

Our complete framework consists of an intensity guidance denoising model and a dual-encoder network for FLIM denoising based on U-Net~\cite{ronneberger2015u}, as illustrated in \cref{fig:architecture}.

\subsubsection{Pre-trained Intensity Model}
The pre-trained intensity model $f_\beta$ employs a three-level U-Net architecture optimized using a hybrid loss function. This function combines a self-supervised Noise2Void loss $\mathcal{L}_{N2V}$~\cite{krull2019noise2void} with a supervised mean squared error (MSE) loss $\mathcal{L}_{MSE}$:
\begin{equation}
   \mathcal{L}_{pretrain} = \alpha \mathcal{L}_{N2V} + (1-\alpha) \mathcal{L}_{MSE}.
\end{equation}
In the encoder pathway, image features are progressively refined through cascaded double-convolution blocks, with the number of feature channels sequentially increased from $1$ to $32$, $64$, $128$, and finally $256$. Each convolutional layer is followed by a LeakyReLU activation to facilitate robust gradient propagation. Symmetrically, the decoder pathway reconstructs spatial details using transposed convolutions, integrating low-level features from the encoder through dense skip connections. These skip connections ensure fine-grained spatial information is retained. The resulting pre-trained model provides denoised intensity estimates $f_\beta(y_I)$, which act as structural priors for the primary FLIM denoising network, enabling the transfer of knowledge from conventional fluorescence microscopy images to enhance FLIM denoising performance.

\subsubsection{Dual-Encoder FLIM Denoising Network}
Our main network processes FLIM channels $y_g$ and $y_s$ through separate encoder paths, each with two hierarchical levels (1→32→64 channels) using $3\times3$ kernels, BatchNorm, and ReLU activations. The network utilizes three specialized decoders: two channel-specific decoders that independently reconstruct $\hat{y}_g$ and $\hat{y}_s$ using features from their respective encoder paths to preserve channel-specific information; and a fusion decoder that produces $\hat{y}_I$ by leveraging features from both encoder pathways through multi-level feature fusion.

The network is initialized with He normal weight initialization to facilitate robust zero-shot capability on each sample. The fusion decoder begins by concatenating the deepest features from both encoders and progressively upsamples through transposed convolutions. At each level, skip connections incorporate encoder features from both pathways:
\begin{equation}
    F_{fusion} = [F_g^L; F_s^L], \quad \hat{F}^l = \mathcal{D}^l(\hat{F}^{l+1}, [F_g^l; F_s^l]),
\end{equation}

where $[;]$ denotes channel-wise concatenation and $\mathcal{D}^l$ represents upsampling operations followed by double convolution blocks that reduce feature dimensions (128→64→32→1).

This design ensures that the fusion decoder incorporates comprehensive latent space information, enabling effective intensity-guided reconstruction from $y_g$ and $y_s$. By strategically combining features from both encoder pathways across multiple scales, the network optimizes information flow from both shallow spatial details and deep semantic features. This integration enhances the ability of decoders for the $g$ and $s$ channels to produce more accurate reconstructions of the FLIM components.
\subsection{Loss Function}
\label{subsec:loss}

Our framework is optimized using a comprehensive joint loss function that try to address spatial resolution and cross-channel consistency of FLIM signal fidelity to restore the images. This approach also preserves cross channel information for lifetime accuracy. The loss function combines multiple terms as shown:
\begin{equation}    \mathcal{L} = \mathcal{L}_{intensity} + \lambda_1 \mathcal{L}_{fidelity} + \lambda_2 \mathcal{L}_{structure} + \lambda_3 \mathcal{L}_{TV}.
\end{equation}

The \textbf{intensity loss} ($\mathcal{L}_{intensity} = \|\hat{y}_I - f_\beta(y_I)\|^2_2$) aligns our predicted intensity with the pre-trained denoising output. This serves as the intensity denoising prior for the model optimization. The \textbf{fidelity loss} ($\mathcal{L}_{fidelity} = \|y_g - \hat{y}_g\|^2_2 + \|y_s - \hat{y}_s\|^2_2$) ensures consistency between inputs and reconstructions, thereby preserving the intrinsic signal characteristics. This component effectively constrains the optimization parameters within physiologically plausible bounds, which is particularly crucial as $y_g$ and $y_s$ encode the lifetime information that should be accurately preserved during the learning process.

The \textbf{structure loss} ($\mathcal{L}_{structure} = [2 - \text{SSIM}(\hat{y}_g, \hat{y}_I) - \text{SSIM}(\hat{y}_s, \hat{y}_I)]$) maintains structural coherence across different reconstruction components. Since the estimated parameters $\hat{y}_s$ and $\hat{y}_g$ represent intensity scaling $g$ and $s$, their structural information should correlate closely with the estimated intensity $\hat{y}_I$ to preserve biologically relevant morphological features in the reconstructed images. Finally, the \textbf{total variation loss} ($\mathcal{L}_{TV} = [\text{TV}(\hat{y}_g) + \text{TV}(\hat{y}_s)]$) reduces noise in relatively uniform areas while preserving sharp edges and preventing over-smoothing of critical details. This regularization term is particularly appropriate due to the normalization constraints inherent in the deconvolution process, effectively limiting the parameter space during optimization and mitigating potential overfitting when gradient steps are excessively large.

By prioritizing intensity and fidelity, and adding structure and total variation constraints, our model delivers effective FLIM denoising that preserves the essential fluorescence lifetime information needed for biological analysis. A comprehensive assessment of these loss components and their relative contributions is detailed in the ablation study presented as in~\cref{subsec:ablation}.

\subsection{Zero-Shot Training Strategy}
\label{subsec:zs-training}
The intensity and lifetime channels influence each other in the noise distribution but do not follow exactly the same distribution because they encode different biological information during measurement. Accordingly, our zero-shot approach directly processes real-world FLIM data containing natural Poisson-Gaussian mixed photon noise statistics in the intensity channel. This method is implemented as a test-time optimization per image, which means it functions in a zero-shot manner without requiring any fine-tuning on new samples. Rather than learning the distribution of the other channel in a supervised way, it instead draws direct inspiration from the intensity channel.

Our implementation employs an iterative optimization procedure using Adam with an initial learning rate of \(1 \times 10^{-3}\) and weight decay of \(1 \times 10^{-5}\). The learning rate is adaptively reduced by a factor of 0.5 when progress plateaus. Each denoising procedure processes randomly cropped \(256 \times 256\) patches from FLIM images, with optimization continuing for $1k$ to $2k$ iterations to ensure convergence. Compared to conventional deep-learning denoising methods that often require lengthy training phases, our zero-shot approach completes optimization in 10--20 seconds on an NVIDIA GeForce RTX 3090 GPU (24GB VRAM) with CUDA acceleration per test. This rapid performance significantly benefits time-sensitive biological imaging tasks by minimizing computational overload while maintaining high-quality FLIM reconstructions.

%% file: sec/4_experiments.tex
\section{Experiments}
\label{sec:experiments}

\begin{figure*}[t]
\centering
\includegraphics[width=0.95\textwidth]{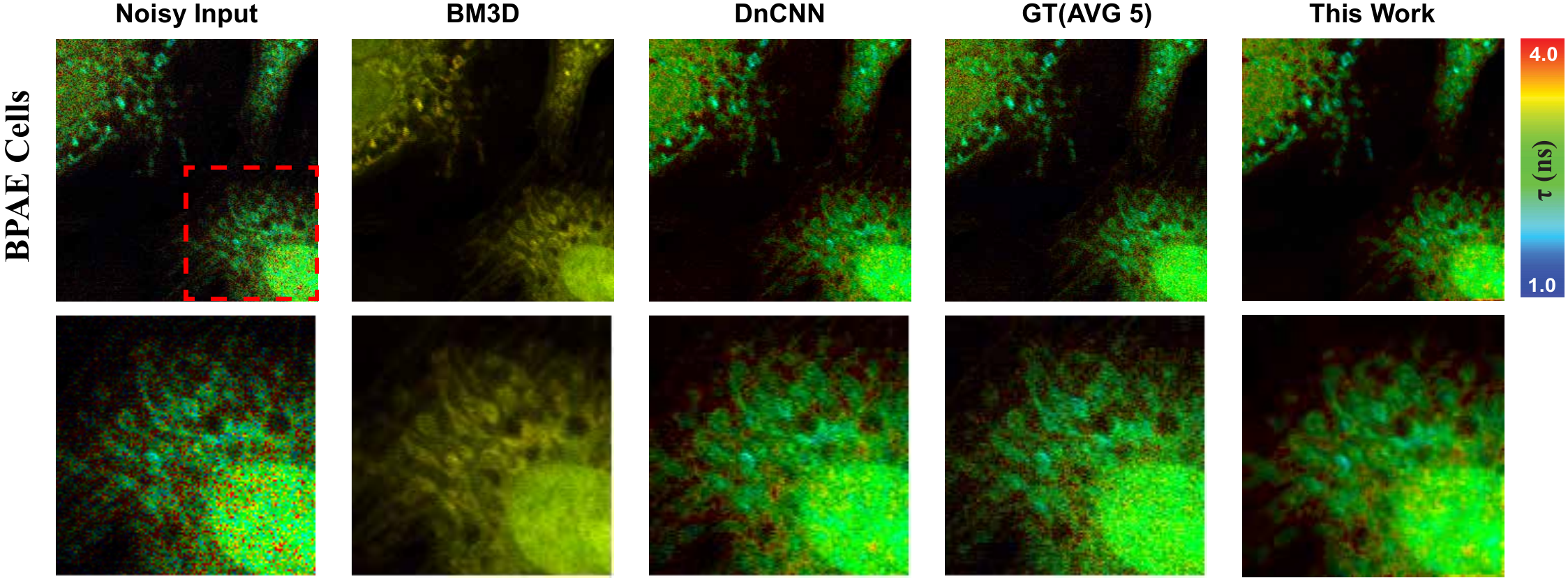}
\caption{Visual comparison of FLIM image denoising methods applied to \textit{Ex vivo} BPAE cells. Top row: full-field images; bottom row: zoomed-in regions corresponding to the highlighted areas. From left to right: Noisy input, BM3D, DnCNN, Ground Truth(AVG 5), and our proposed zero-shot method. The color bar represents fluorescence lifetime values ranging from 1--4 ns. the proposed method in this work demonstrates superior preservation of lifetime information comparable to AVG 5 while achieving effective noise reduction.}
\label{fig:results}
\end{figure*}
We evaluate the proposed Zero-Shot Denoising Framework for FLIM using empirical data in this section. Our methodology incorporates diverse regions of interest (ROI) across multiple specimens to capture authentic noise variations representative of real-world imaging conditions. Through comprehensive experiments across diverse biological specimens---including \textit{In vivo}, \textit{Ex vivo}, and fixed samples with different fluorophores---we assess the efficacy of our approach against state-of-the-art denoising methods on real-world FLIM data.

Given the correlation and distinct distribution patterns across channels of FLIM data, coupled with the inherent limitations in dataset availability, we concentrate our evaluation on two critical aspects: (1) the quality of reconstructed lifetime images and their corresponding $g$ and $s$ components, and (2) the accuracy of lifetime estimations produced by our framework. Our comprehensive analysis demonstrates significant improvements over existing state-of-the-art methods in both qualitative and quantitative metrics. We also conduct comprehensive ablation studies to measure how each loss function uniquely contributes to the overall performance of the model. These experiments reveal important insights about how our different optimization objectives work together.

\subsection{Experimental Setup}
\label{subsec:setup}

\subsubsection{Model Pretraining}
\label{subsubsec:pre-trained}
Prior to application on FLIM data denoising, we apply a pre-trained U-Net architecture backbone network on real-world fluorescence microscopy images containing mixed Poisson-Gaussian noise \cite{zhang2019poisson}. This pretraining strategy enables our model to learn the statistical properties of noise distributions common in fluorescence imaging without requiring paired clean-noisy training data specific to FLIM. By utilizing general fluorescence microscopy data for pretraining rather than FLIM-specific images, our network develops generalizable denoising principles that can be effectively transferred to FLIM data during inference through our zero-shot framework. This strategy helps overcome the limited availability of matched noisy-clean FLIM image pairs for training.

\subsubsection{Data Acquisition}
\label{subsec:data_acquisition}
Due to the limited availability of public FLIM data, we evaluated the proposed method on five distinct biological specimens from both published datasets and newly acquired images with two-photon excitation at an 800 nm wavelength by ``Instant-FLIM''~\cite{Zhang21Instant}. These specimens, spanning a range of sample types and fluorescence characteristics, were employed for quantitative evaluation: (1) \textbf{\textit{In vivo} Zebrafish}: labeled with enhanced green fluorescent protein (EGFP)~\cite{Zhang21Instant}; (2) \textbf{\textit{Ex vivo} Mouse Kidney}: tissue sections were stained with DAPI (nuclei), Alexa Fluor 488 wheat germ agglutinin (cell membranes), and Alexa Fluor 568 phalloidin (F-actin) to visualize key structural components~\cite{mannam2023improving}; (3) \textbf{\textit{Ex vivo} BPAE Cells}: bovine pulmonary artery endothelial cells were labeled with DAPI (nuclei), Alexa Fluor 488 phalloidin (F-actin), and MitoTracker Red CMXRos (mitochondria) to enable multicomponent lifetime analysis~\cite{Zhang19seg}; (4) \textbf{Fixed NIH3T3 Cells}: fibroblasts were stained with DAPI (nuclei), Alexa Fluor 488 phalloidin (F-actin), and Alexa Fluor 594 biocytin (AFB) to examine subcellular structures; and (5) \textbf{\textit{In vivo} Mouse Kidney}: intrinsic autofluorescence from kidney tissue was imaged in live animal models to assess performance under physiological conditions~\cite{mannam2020machine,Zhang19seg}. 

Due to the lack of clean benchmarks for (5) \textit{In vivo} kidney, we only compare this result to DnCNN~\cite{mannam2023improving}. This diverse dataset represents a comprehensive testbed for fluorescence lifetime image denoising algorithms, encompassing both exogenous fluorophores with targeted subcellular localization and endogenous autofluorescence signals across different tissue types. 

\subsection{Real-World Noise Reduction}
\label{subsec:real-world}

The ground truth was established by averaging 5-15 images (AVG 5-15) for each FLIM component on the same ROIs, carefully balancing signal quality with the need to minimize photobleaching and photodamage to biological samples during extended imaging sessions. For each denoising method, we calculated the peak signal-to-noise ratio (PSNR) and structural similarity index (SSIM) for both $g$ and $s$ components. Additionally, we computed the Absolute Lifetime Error (ALE), which quantifies the percentage deviation of calculated fluorescence lifetime values from the ground truth. Comparative results across all samples with averaged result of $g$ and $s$ are summarized in \cref{tab:comparison}.

The quantitative evaluation demonstrates that our proposed approach consistently outperforms both traditional (BM3D) and pre-trained deep learning (DnCNN) methods by maintaining important FLIM lifetime features while effectively reducing noise across various biological samples. The proposed method notably achieves higher PSNR and SSIM values compared to pre-trained DnCNN and BM3D, highlighting its enhanced capacity for accurate signal recovery. Critically, our technique significantly reduces the ALE, thus providing more precise fluorescence lifetime estimations crucial for reliable biological interpretations. These improvements underscore the efficacy and robustness in practical FLIM imaging scenarios.

Additionally, qualitative analysis of lifetime images with HSV color mapping (lifetime as Hue and intensity as Value) reveals significant performance differences across methods. As shown in \cref{fig:results}, which displays cellular structures, BM3D partially denoises individual $g$ and $s$ components but significantly distorts lifetime values. Similarly, DnCNN fails to preserve critical information in nuclei regions despite its pre-trained intensity channel denoising, as it cannot account for channel correlation information. The proposed method, however, effectively preserves morphological details while reducing noise. 

\begin{figure}
\centering
\includegraphics[width=\columnwidth]{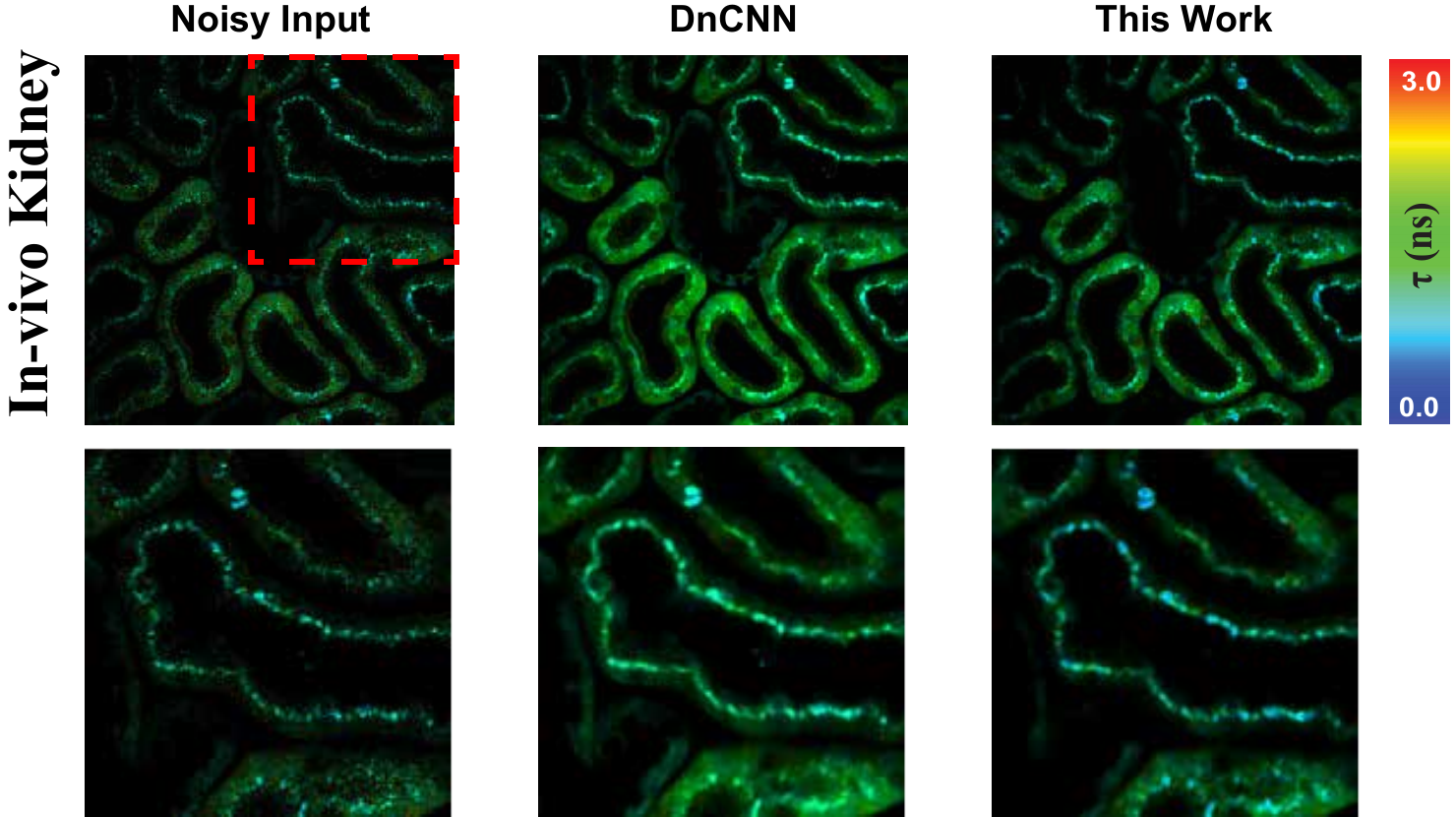}
\caption{Visual comparison of FLIM image denoising methods for \textit{In vivo} Mouse Kidney. Top row: full-field images; bottom row: zoomed-in regions corresponding to the highlighted areas. From left to right: Noisy input, pre-trained DnCNN, and our proposed method. The color bar represents fluorescence lifetime values ranging from 0--3 ns.}
\label{fig:results2}
\end{figure}

\cref{fig:results2} provides further evidence of the superiority from this work through \textit{in vivo} Mouse Kidney imaging. Our approach maintains accurate fluorescence lifetime values across different tissue regions, whereas DnCNN introduces visible bias. Our zero-shot denoising performs consistently across diverse biological samples, highlighting its robustness. By processing all channels simultaneously and leveraging pre-trained intensity maps, our approach effectively preserves essential lifetime information while eliminating noise.

\begin{table}
\centering
\caption{Quantitative comparison across biological specimens. Best results are in \textbf{bold}.}
\label{tab:comparison}
\begin{tabular}{@{}lccc@{}}
\toprule
\textbf{Sample} & \multicolumn{3}{c}{\textbf{PSNR (dB)}} \\
\cmidrule{2-4}
 & BM3D & DnCNN & \textbf{This work} \\
\midrule
\textit{In vivo} Zebrafish & 27.31 & 34.25 & \textbf{37.22} \\
\textit{Ex vivo} Mouse Kidney & 41.56 & 46.91 & \textbf{54.87} \\
\textit{Ex vivo} BPAE & 34.03 & 38.81 & \textbf{39.13} \\
Fixed NIH3T3 & 3.30 & 23.41 & \textbf{39.17} \\
\midrule
\textbf{Average} & 26.55 & 35.84 & \textbf{42.60} \\
\midrule
\textbf{Sample} & \multicolumn{3}{c}{\textbf{SSIM}} \\
\cmidrule{2-4}
 & BM3D & DnCNN & \textbf{This work} \\
\midrule
\textit{In vivo} Zebrafish & 0.4100 & 0.7379 & \textbf{0.8562} \\
\textit{Ex vivo} Mouse Kidney & 0.8450 & 0.9948 & \textbf{0.9981} \\
\textit{Ex vivo} BPAE & 0.7053 & 0.7526 & \textbf{0.7698} \\
Fixed NIH3T3 & 0.4016 & 0.6207 & \textbf{0.9397} \\
\midrule
\textbf{Average} & 0.5905 & 0.7765 & \textbf{0.8909} \\
\bottomrule
\textbf{Sample} & \multicolumn{3}{c}{\textbf{ALE (\%)}} \\
\cmidrule{2-4}
 & BM3D & DnCNN & \textbf{This work} \\
\midrule
\textit{In vivo} Zebrafish & 60.99 & 103.93 & \textbf{56.41} \\
\textit{Ex vivo} Mouse Kidney & 23.59 & 11.73 & \textbf{10.81} \\
\textit{Ex vivo} BPAE & 73.48 & 57.21 & \textbf{52.60} \\
Fixed NIH3T3 & 23.93 & 6.09 & \textbf{5.44} \\
\midrule
\textbf{Average} & 45.50 & 44.74 & \textbf{31.31} \\
\bottomrule
\end{tabular}
\end{table}

\begin{figure*}
\centering
\includegraphics[width=0.96\textwidth]{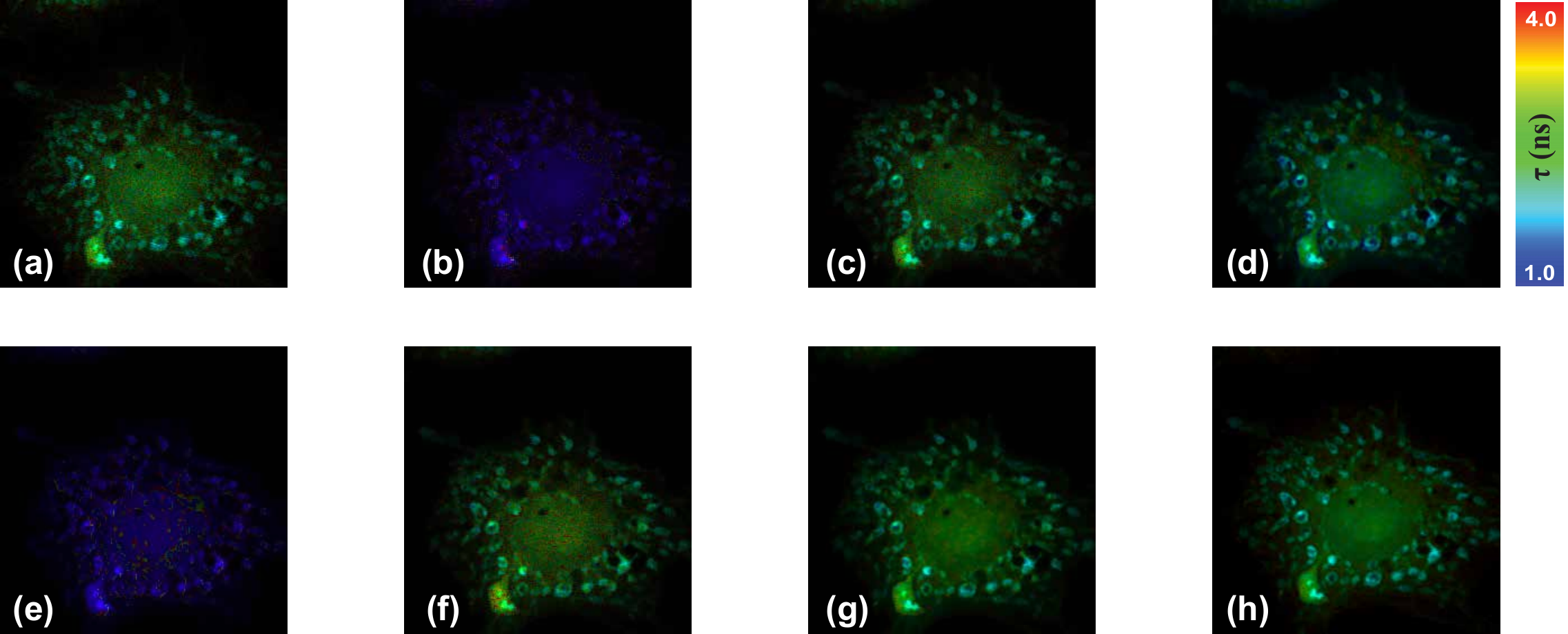}
\caption{Ablation study showing the impact of loss components on lifetime image denoising of \textit{Ex vivo} BPAE cells. (a) Noisy input. (b) Using only $\mathcal{L}_{intensity}$. (c) Adding $\mathcal{L}_{fidelity}$. (d) Adding $\mathcal{L}_{structure}$. (e) Adding $\mathcal{L}_{TV}$. (f) Adding $\mathcal{L}_{fidelity}+\mathcal{L}_{structure}$. (g) Full loss function. (h) Ground truth reference (AVG 15).}
\label{fig:ablation}
\end{figure*}

\subsection{Ablation Studies}
\label{subsec:ablation}
We evaluated the contribution of individual components in our proposed loss function (\cref{subsec:loss}) to the overall performance of our FLIM denoising framework. We conducted a series of ablation experiments by systematically removing or altering specific terms in our composite loss function and quantifying the resulting impact.
\begin{table}
\centering
\caption{Ablation study results comparing different loss function combinations. We measured PSNR (dB), SSIM, and Average Lifetime Error (ALE, \%).}
\label{tab:ablation}
\small
\begin{tabular}{@{}lccc@{}}
\toprule
\textbf{Loss Configuration} & \textbf{PSNR $\uparrow$} & \textbf{SSIM $\uparrow$} & \textbf{ALE $\downarrow$} \\
\midrule
$\mathcal{L}_{intensity}$ & 33.06 & 0.5841 & 414.62 \\
\quad $+ \lambda_1 \mathcal{L}_{fidelity}$ & 41.34 & 0.8600 & 55.98 \\
\quad $+ \lambda_2 \mathcal{L}_{structure}$ & 33.29 & 0.7258 & 46.91 \\
\quad $+ \lambda_3 \mathcal{L}_{TV}$ & 34.23 & 0.6180 & 112.17 \\
\quad $+ \lambda_1 \mathcal{L}_{fidelity} + \lambda_2 \mathcal{L}_{structure}$ & 41.41 & 0.8637 & 51.94 \\
\textbf{All Loss} & \textbf{41.85} & \textbf{0.8810} & \textbf{34.09} \\
\bottomrule
\end{tabular}
\end{table}

As presented in~\cref{tab:ablation}, each loss component contributes differently to FLIM denoising performance across all evaluated samples. The visual impact of these components is illustrated in~\cref{fig:ablation} using the BPAE cell sample as a representative example. The intensity loss ($\mathcal{L}_{intensity}$) alone effectively removes noise and achieves reasonable structural quality. However, due to the global optimization approach, it struggles to accurately preserve fluorescence lifetime information, resulting in a high ALE. Incorporating the fidelity loss ($\mathcal{L}_{fidelity}$) significantly enhances structural accuracy and lifetime consistency by enforcing signal-level coherence between input and reconstruction, although it insufficiently addresses noise reduction in the lifetime channels, as specifically demonstrated in the BPAE cell sample shown in~\cref{fig:ablation}(c).

The structure loss ($\mathcal{L}_{\text{structure}}$) enhances structural coherence and significantly reduces ALE; however, it introduces undesirable shifts in lifetime values as observed in~\cref{fig:ablation}(d), particularly in mitochondrial staining for the BPAE sample. Similarly, total variation loss ($\mathcal{L}_{TV}$) encourages smoothness while preserving edges, which helps maintain detailed structures. However, when used alone, it damages both structural integrity and lifetime accuracy, as shown in~\cref{fig:ablation}(e).

Combining fidelity and structure losses (see~\cref{fig:ablation}(f)) emphasizes signal fidelity but remains insufficient for effective noise suppression. Only our full combined loss formulation (\cref{fig:ablation}(g)) achieves an optimal balance, effectively suppressing noise while accurately preserving the fluorescence lifetime characteristics. These results clearly validate the necessity and effectiveness of our comprehensive loss function design. Without the intensity guidance, the other loss components individually during the optimization process fail to return meaningful information, potentially causing the solution to diverge from the convex optimization path.

The ablation results confirm the complementary effectiveness of our carefully designed loss components. The complete loss function consistently outperforms all partial variants across all evaluated metrics, achieving substantial improvements in both image quality (PSNR, SSIM) and lifetime accuracy (ALE).

\subsection{Discussion}
\label{subsec:discussion}
The efficacy of our framework is dependent on the intensity prior, with enhanced performance directly yielding improved results across all lifetime channels. While our framework currently demonstrates excellence in denoising fluorescence lifetime microscopy, its underlying methodology extends beyond this specific application. The proposed architecture can be seamlessly adapted to other multimodal microscopy techniques with similar correlated noise patterns.

Notably, our approach offers promising capabilities for super-resolution microscopy and deconvolution tasks by providing a computationally efficient solution with exceptional performance. This method requires no training data ---a significant advantage in microscopy domains where annotated datasets are often limited.

By addressing the fundamental challenges of noise, resolution, and blur through a unified methodological approach, our framework represents a significant advancement in microscopy image enhancement technology with potential impact across multiple imaging modalities. These capabilities establish our approach as highly applicable to quantitative biomedical imaging, where precise downstream analyses depend critically on high-quality input data.

%% file: sec/5_conclusions.tex
\section{Conclusions}
\label{sec:conclusion}
In this paper, we introduced a novel Zero-Shot Denoising Framework for FLIM with an Intensity-Guided Learning approach that effectively addresses the fundamental challenge of denoising multi-channel data in FLIM without requiring training data for additional channels. By connecting fluorescence intensity and lifetime channels, our network significantly reduces noise while preserving critical lifetime information. The comprehensive loss function design ensures effective denoising while maintaining the integrity of information across all channels. Experimental results on real-world datasets demonstrate that the proposed approach enables more reliable extraction of physiological and molecular information from challenging multimodal data acquisitions and potentially expands the applicability of this powerful technique to other related imaging techniques.